\documentclass[useAMS,usegraphicx]{mn2e}
\usepackage{rotate}
\usepackage{times}
\newif\ifAMStwofonts
\AMStwofontstrue

\def\gs{\mathrel{\hbox{\rlap{\hbox{\lower4pt\hbox{$\sim$}}}\hbox{$>$}}}}
\def\ls{\mathrel{\hbox{\rlap{\hbox{\lower4pt\hbox{$\sim$}}}\hbox{$<$}}}}

\def\Msun{M$_{\odot}$}

\def\xmm{{\it XMM-Newton}}

\def\xte{{\it RXTE}}
\def\xmm{{\it XMM-Newton}}

\def\xeus{{\it XEUS}}

\def\et{{et al.\ }}

\def\Msun{\hbox{$\rm\thinspace M_{\odot}$}}

\title[Where are the QPOs in AGN? ]
      {Where are the X-ray QPOs in active galaxies?}

\author[S. Vaughan \& P. Uttley]
       {S. Vaughan$^{1}$ and
        P. Uttley$^{2}$ \\
$^{1}$X-Ray and Observational Astronomy Group, University of
       Leicester, Leicester, LE1 7RH\\
$^{2}$X-ray Astrophysics Laboratory, Code 662,
NASA Goddard Space Flight Center, Greenbelt Road, Greenbelt, MD 20771, USA
}
\date{Accepted 16/06/2005; submitted 23/05/2005; in original form 08/04/2005}
\pagerange{\pageref{firstpage}--\pageref{lastpage}}
\pubyear{2005}
\begin{document}
\maketitle
\label{firstpage}

\begin{abstract}
In this paper  we address the question of whether existing X-ray
observations of Seyfert galaxies are sufficiently sensitive to detect
quasi-periodic oscillations (QPOs) similar to those observed in the
X-ray  variations of Galactic Black Holes (GBHs). We use data from
\xmm\ and simulated data based on the best \xte\ long-term monitoring
light curves, to show that if X-ray QPOs are present in Seyfert X-ray
light curves -- with similar shapes and strengths to those observed in
GBHs, but at lower frequencies commensurate with their larger black
hole masses -- they would be exceedingly difficult to detect.  Our
results offer a simple explanation for the present lack of QPO
detections in Seyferts. We discuss the improvements in telescope size
and monitoring patterns needed to make QPO detections feasible.  The
most efficient type of future observatory for searching for X-ray QPOs
in AGN is an X-ray All-Sky Monitor (ASM). A sufficiently sensitive ASM
would be ideally suited to detecting low frequency QPOs in nearby
AGN. The detection of AGN QPOs would strengthen the AGN-GBH connection
and could serve as powerful diagnostics of the black hole mass, and
the structure of the X-ray emitting region in AGN.
\end{abstract}

\begin{keywords}
galaxies: active -- galaxies: Seyfert: general -- X-ray: galaxies  
\end{keywords}


\section{Introduction}
\label{sect:intro}

One of the most powerful approaches to studying Galactic Black Holes
(GBHs), and X-ray binaries (XRBs) in general, is through X-ray timing. See van
der Klis (1995, 2005) and M$^{\rm c}$Clintock \& Remillard (2005) for
detailed reviews. Perhaps the most widely used tool of time series
analysis is the power spectrum, sometimes known as the
power spectral density (PSD; see van der Klis 1989a 
for a thorough review of power spectral methods as applied to XRBs).  

The power
spectra of XRBs are usually described in terms of  two types of
variability  The first is aperiodic variability, usually called
`noise,' which is often band-limited, meaning that the power spectrum
is a broad continuum over a wide (but finite) frequency range.  The
second type of variation are quasi-periodic oscillations (QPOs). These
represent a concentration of variability power over a limited
frequency range, usually revealed by a well-resolved peak in the power
spectrum\footnote{This type of variation should not be confused with
strictly   periodic oscillations, in which all the power in
concentrated entirely at specific frequencies resulting in unresolved
peaks in power spectra. Strictly periodic variations have been
observed in neutron star XRBs, but not GBHs.}. QPOs are one of the
most powerful diagnostics of XRBs: they depend on the evolution of the
source through different accretion states and the highest frequency
QPOs are thought to be tracers of matter orbiting in the strongly
curved spacetime around the black hole.

For many years it has been known that the radio-quiet Active Galactic
Nuclei (AGN), and Seyfert galaxies in particular, show substantial
X-ray variability (e.g. Barr \& Mushotzky 1986; Lawrence \et 1987).
For almost as many years the X-ray variability characteristics of AGN have been
compared to those of the better-understood GBHs (e.g. M$^{\rm c}$Hardy 1988). 
This comparison is based on the
premise that the physics of accretion on to stellar-mass black holes
($M_{\rm BH} \sim 10$~\Msun) should be essentially the
same for accreting supermassive black holes ($M_{\rm BH} \sim
10^6 - 10^9$~\Msun) that power AGN. Indeed, it has now been 
shown that the X-ray variations of Seyfert galaxies have noise
power spectra that are remarkably similar to those of GBHs (see e.g. 
Edelson \& Nandra 1999; Uttley, M$^{\rm c}$Hardy \& Papadakis 2002; Markowitz \et 2003;
Vaughan, Fabian \& Nandra 2003; M$^{\rm c}$Hardy et al. 2004, 2005). 
The available data are broadly
consistent with the hypothesis that Seyfert galaxies possess the same noise power
spectra as GBHs, but with much lower characteristic frequencies due to
these frequencies scaling with $M_{\rm BH}^{-1}$, i.e. the
characteristic variability timescales depend linearly on the mass
(size) of the central black hole. Given this
connection between the noise spectra of Seyferts and GBHs, it seems
reasonable to suppose Seyfert galaxies might also possess X-ray QPOs
similar to those of GBHs,
albeit at accordingly lower frequencies. Curiously, despite
improvements in observing sensitivity and scheduling flexibility,
there are still no robust detections of X-ray QPOs in Seyferts,
notwithstanding occasional claims to the contrary (see Benlloch \et
2001; Vaughan 2005 and Vaughan \& Uttley 2005 for discussions).

The purpose of this paper is to answer the following question:
is the lack of X-ray
QPO detections in Seyferts due to insufficient observational
sensitivity or a genuine absence of QPOs in Seyferts? The latter would
indicate an important difference between supermassive and stellar mass
accreting  black holes.  The plan of the paper is as
follows.  Section~\ref{sect:qpo} gives a very brief review of 
the two main QPO types commonly observed in GBHs and how these might appear
if present in Seyfert galaxies.   Section~\ref{sect:hfqpo} uses the
best available \xmm\ observations of Seyferts to search for high
frequency QPOs, and section~\ref{sect:lfqpo} discusses whether
long-term \xte\ monitoring  is capable of detecting low frequency
QPOs.  Finally, section~\ref{sect:disco} argues that the lack of a
robust, positive detection of an X-ray QPO in a Seyfert is not
inconsistent with the idea that at least some AGN possess QPOs similar to those
of GBHs, and describes the observational improvements needed for
more sensitive QPO searches.

\begin{figure*}
\centering
\includegraphics[width=6.0 cm, angle=-90]{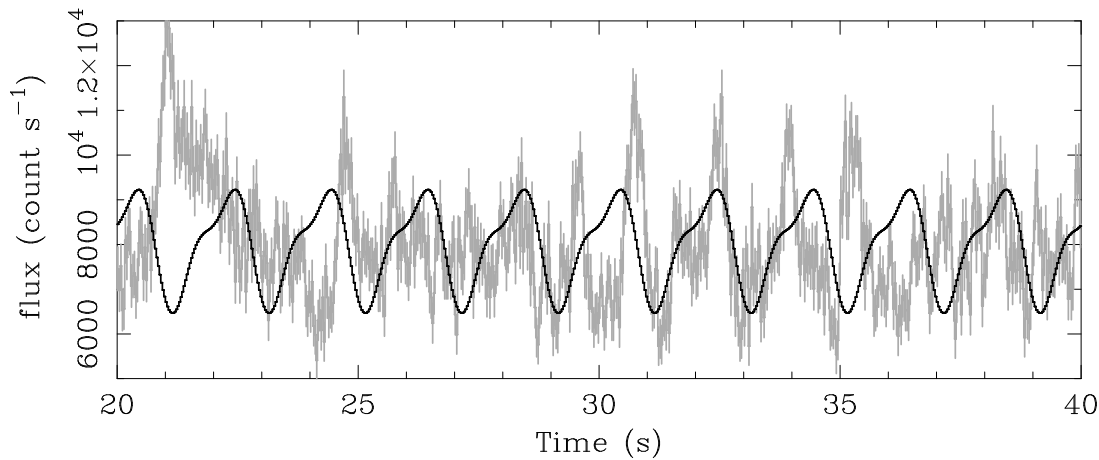}
\caption{
\xte\ light curve of GRS 1915+105 in the $2-13$~keV band
while the source showed a strong $f_0 \simeq 0.5$~Hz QPO
(and a $1$~Hz harmonic). 
The black line shows the strictly periodic double-sinusoid
model fitted to the cycle between $31$ and $33$~s and extrapolated
to the rest of the light curve, demonstrating clearly that with short
data segments even such a strong and coherent QPO signal
is difficult to discern.
\label{fig:qpo_lc}}
\end{figure*}


\section{What should we expect for AGN QPOs?}
\label{sect:qpo}

From an observational perspective QPOs are loosely defined as
resolved peaks in the PSD (van der Klis 1989b). 
The QPOs commonly observed in GBHs are often
described by a Lorentzian profile in the PSD,
\begin{equation}
L(f) = \frac{ R^2 Q f_0 / \pi }{ f_0^2 + Q^2 (f - f_0)^2}
\end{equation}
where $f_0$ is the frequency of the Lorentzian, $Q$ is the quality
factor that determines its width ($Q \approx f_0/\Delta f$) and $R$ is
a normalisation factor giving the total strength of the QPO (for high
$Q$ values $R$ is approximately equal to the fractional rms amplitude
of variability in the QPO).  $Q$
essentially denotes the coherence of the variation, becoming infinite in
the strictly periodic case, though by definition this is never realised for 
QPOs.  The distinction between a QPO and
band limited noise becomes unclear for very broad profiles ($Q \ls
2$), perhaps indicating  
a common physical origin for these features (e.g. Nowak 2000;
Belloni, Psaltis \& van der Klis 2002). However, for the purpose of
the present discussion, QPOs shall be defined as those PSD
components with $Q \gs 3$ (i.e. relatively coherent oscillations).

In simple terms, the Lorentzian profile is the PSD of an
exponentially decaying sinusoid.  However, the Lorentzian shape can be
produced by a variety of different mathematical models, for example
by a single oscillation that is
perturbed in frequency, amplitude or phase, or through
filtering a purely stochastic process (a broad-band noise spectrum) to
suppress variability power below and above the peak frequency.
Naturally, this insensitivity to the exact form of the light curves
means that there is considerable scope for different physical models
to produce QPOs; higher order statistics (beyond the power
spectrum) may be required to distinguish
between specific models for the origin of QPOs (e.g. Maccarone \& Schnittman
2005).   It is therefore simplest to use the empirical phenomenology
of GBH QPOs to infer what might be expected to be observed in AGN, rather
than assume a specific physical model of their origin.

There are two broad classes of QPO observed in GBHs: low frequency
QPOs (LF QPOs) in the range $\sim 50$~mHz to $\sim 30$~Hz, and high frequency
QPOs (HF QPOs) usually at frequencies $\gs 100$~Hz. 
HF QPOs are the highest-frequency variations yet observed from GBHs,
with quasi-periods similar to the expected dynamical time-scales in the inner
accretion disk. There is tentative evidence that HF QPOs occur
at specific, fixed 
frequencies for each source, perhaps 
related to the black hole mass (Abramowicz \et 2004; M$^{\rm
c}$Clintock \& Remillard 2005). They are relatively weak variations,
with $R \sim 1 - 5$ per cent rms, and $Q$ values ranging from $3-20$ (see
e.g. Remillard \et 2002a). HF QPOs seem to be preferentially detected
in the very high states of GBHs.

By contrast, the LF QPOs can
be quite narrow ($Q>5$) and strong (with $R \sim 3 - 15$ per cent
rms) and have been observed in both the low/hard and intermediate/very
high states of GBHs.  Such well-defined, strong peaks are much less
common in the high/soft state.  In any given source the frequencies of LF 
QPOs can vary substantially (but on time-scales of hours or more, which
correspond to centuries or even longer for AGN).
Because of this variability, similar LF QPOs
can be observed over at least a decade range in frequency.  In this paper
only the strongest LF QPOs observed in GBHs, which are
most likely to be detectable in AGN, will be considered. Figure~\ref{fig:qpo_lc}
shows a short segment of an X-ray light curve
of GRS 1915+105, a GBH, which showed a strong QPO at $f_0 = 0.5$~Hz,
$R \simeq 14$~per~cent and $Q \simeq 7$.
The observation was made on 1996 Jul 23. The QPO also shows
a weaker harmonic at $1$~Hz. This QPO was modelled in the time domain
by fitting a sum of sinusoids to the single `cycle' between
$31$ and $33$~s on the time axis, and extrapolating the sinusoidal model
to lower and higher times for comparison with the data.  
It is clear from the figure that the variations are neither
strictly periodic nor do they show a well-defined profile.
Even with this very strong, coherent QPO, observation of
many `cycles' is required before the quasi-periodicity can be distinguished from
the aperiodic signal (i.e. the underlying band-limited noise). 

If Seyfert galaxies (and AGN in general) are the  supermassive
analogues of GBHs, and show similar broad-band noise variability in
X-rays, it seems plausible that Seyferts should also show similar QPOs
to those seen in GBHs. The simplest hypothesis is that the QPOs retain
the same relative strengths and shapes (i.e. same $R$ and $Q$) as seen
in GBHs, but with greatly reduced frequencies due to the inverse
scaling of frequency with black hole mass.  For LF QPO frequencies of
order $\sim1$~Hz in GBHs, assuming a typical black hole mass of $\sim
10$~\Msun, the analogous LF QPOs in a Seyfert galaxy should occur
around $f_{\rm LFQPO} \sim 10^{-5} (M_{\rm BH}/10^6 \Msun)^{-1}$~Hz
(i.e. timescales of $\gs 100$~ks).  For HF QPOs, the $450$~Hz QPO in
GRO J1655-40 (Strohmayer 2001) was taken as being representative of
the class.  Assuming a black hole mass of $\sim 6$~\Msun\ (Shahbaz \et
1999), the inverse scaling between frequency and black hole mass
yields an expected frequency of $f_{\rm HFQPO} \sim 3 \times 10^{-3}
(M_{\rm BH}/ 10^6 \Msun)^{-1}$~Hz (i.e. timescales $\gs 400$~s) for
the analogous HF QPO in Seyfert galaxies.

The other important point to consider when comparing AGN and GBHs is
the energy range involved. QPOs tend to have relatively hard spectra
in GBHs, i.e. the relative amplitude of the QPO increases with energy
through the $2-20$ keV band (e.g. Belloni \et 1997; Tomsick \& Kaaret
2001). 
As no QPOs have been detected in AGN, the energy of the peak amplitude
of the QPOs (if they exist) is unknown, but, assuming QPOs are associated with
the non-thermal X-ray emission component they should occur within the bandpass
of current X-ray missions (e.g. $0.2-10$~keV for \xmm\ and
$3-15$~keV for \xte) since over these energy ranges AGN spectra are dominated
by the non-thermal component. See Done \& Gierlinski (2005) for a discussion
of some other effects of bandpass on the comparison between GBHs and AGN.

In the following two sections
existing and simulated data are used to test whether these LF and HF
QPOs could be detected if they are present in the Seyfert galaxies
which have been most intensively observed in X-rays.


\section{Can we detect high frequency QPOs?}
\label{sect:hfqpo}

According to the arguments given above, a low mass Seyfert galaxy
($M_{\rm BH} \sim 10^6$~\Msun) could be expected to show a HF QPO at
mHz frequencies.
These are accessible using \xmm\ long-look observations which
span $\sim 100$~ks with uninterrupted sampling and high
signal-to-noise, and constrain the variability on timescales as
short as $10-100$~s. The \xmm\ long-looks reliably constrain
the PSD over a range of $\sim 10^{-4} - 10^{-1}$~Hz.

In order to test whether a HF QPO could be present in
existing Seyfert data, the
three orbit \xmm\ observation of the bright  Seyfert 1 galaxy
MCG$-$6-30-15 was extracted  from the public archive. These data
represent the longest \xmm\ observation of a 
Seyfert galaxy performed to date. The PSD of these data
was originally presented by Vaughan \et (2003).  For the present
analysis a background-subtracted light curve of the source was
produced in the $0.2-10$~keV band with $10$-s time bins. 

A periodogram (the modulus squared of the Fourier transform) 
was computed for each orbit of data. The three orbits were then 
combined and the resulting periodogram was averaged over frequency
bins to produce a power spectrum estimate. 
The data were normalised to units of $[{\rm rms/mean}]^2~{\rm Hz}^{-1}$ (see van der
Klis 1997). This normalisation is such that integrating the PSD
between two frequencies gives the variance, relative to the mean flux,
within that frequency range.

\begin{figure}
\centering
\includegraphics[width=5.4 cm, angle=270]{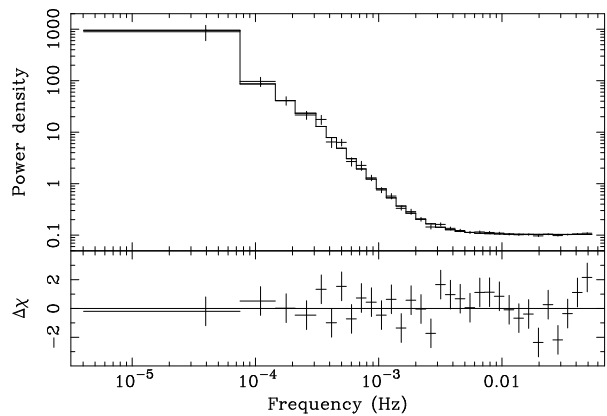}
\caption{
PSD of MCG$-$6-30-15 calculated using the $0.2-10$~keV
light curve extracted from the \xmm\ EPIC pn.
The PSD is in units of $[{\rm rms/mean}]^2~\rm{Hz}^{-1}$.
{\bf The data are shown with crosses and the histogram represents the best
fitting model 
comprising a broken power law (intrinsic broad band noise)
plus a constant (due to instrumental Poisson noise).
The lower panel shows the $(data-model)/\sigma$ residuals. 
}
\label{fig:mcg6psd}}
\end{figure}

The binning was logarithmic in frequency, meaning that bins cover a
range $f \rightarrow f+df$ where $df/f$ is a constant that defines the
resolution of the data.  [The frequency bins are therefore equally
spaced in logarithmic frequency by $\log (1+df/f)$.] The only other
constraint is that  a minimum of $20$ periodogram points were included
in each  bin in order to obtain reasonable estimates of the mean power
and its uncertainty (see Papadakis \& Lawrence 1993).  The fractional
resolution of the binned data is therefore kept roughly constant,
meaning that a QPO of fixed width, $Q$, will occupy the same number of
bins over the frequency range of the PSD.  The search for QPOs is made more
sensitive by binning the data so  that the frequency resolution
matches the expected width of the QPO being sought (see section 5.3 of
van der Klis 1989).  In the present analysis the binning was optimised
to search for  QPOs with $Q = 5$ (i.e. $df/f = 0.2$) which seems to be
typical for GBHs (Remillard \et 2002a,b). The resulting PSD is shown
in Fig.~\ref{fig:mcg6psd}.

As reported by Uttley \et (2002), Vaughan \et (2003) and most recently
(with new long-term monitoring data) by M$^{\rm c}$Hardy \et (2005),
the PSD of MCG$-$6-30-15 is adequately explained by a simple broken
(or bending) power law model. A broken power law was fitted by
adjusting the parameters of the model to  minimise the $\chi^2$
statistic. For the present analysis the best fit was obtained using
standard $\chi^{2}$ minimisation of an analytical model, rather than
through Monte Carlo simulations (see Uttley \et 2002; Vaughan \et
2003).  This latter approach more fully accounts for distortions on
the  continuum PSD due to spectral leakage effects, which can alter
the continuum slope (aliasing is not a problem for the \xmm\
long-looks because the data are contiguously sampled).  The much
faster method of direct fitting of an analytical model was used
in the present analysis since  the sampling distortions should not
substantially affect the detectability of HF QPOs.  The broken
power law model (plus a constant Poisson noise background level)
provided a good fit to the data ($\chi^2 = 37.11$ for $\nu = 27$
degrees of freedom, dof) with a break timescale consistent with that
previously measured by Vaughan \et (2003).  Both an exponentially
cut-off power law and a smoothly bending power law (M$^{\rm c}$Hardy
\et 2004) provided worse fits.  

\begin{figure}
\centering
\includegraphics[width=6.40 cm, angle=270]{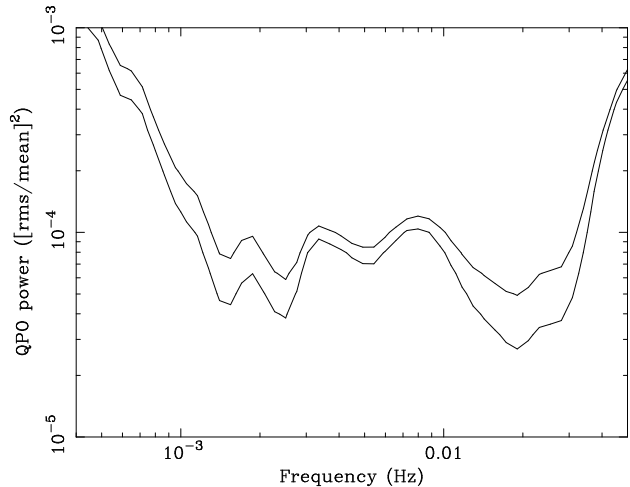}
\caption{
Joint confidence region for QPO frequency and strength ($R^2$) mapped
out using $\Delta \chi^2 = 4.61, 9.21$ contours, based on fitting the PSD
of MCG$-$6-30-15 (see Fig.~\ref{fig:mcg6psd}). 
Given the black hole mass of $1-6 \times 10^6$~\Msun\
a HF QPO might be expected around $f_0 \sim 0.5-3\times 10^{-3}$~Hz.
\label{fig:mcg6qpo}}
\end{figure}

The simple broken power law model provided a satisfactory fit to the
data and, significantly, there were
no obvious systematic residuals to indicate the presence of a QPO
(bottom panel of Fig.~\ref{fig:mcg6psd}).  Therefore, in order to constrain the
strength of any possible weak QPOs a Lorentzian was included
in the model with a quality factor $Q=5$.
Figure~\ref{fig:mcg6qpo} shows the $\Delta \chi^2 = 4.61, 9.21$
contours for QPO strength ($R^2$) against frequency, over 
the useful bandpass of the
data ($f_0 = 4\times 10^{-4} - 5 \times 10^{-2}$~Hz).
These $\Delta \chi^2$ values represent approximate $90$ and $99$
per cent joint confidence regions on the frequency and power of a QPO.
The contours show that a QPO above a few mHz is constrained to have 
$R^2 \ls 10^{-4}$ (corresponding to a total QPO strength of $R \ls
1$ per cent rms). 

Vaughan \et (2003) estimated a mass for MCG$-$6-30-15
of $\sim 10^6$~\Msun\ based on comparing the PSD
break frequency to that of Cyg X-1. This estimate was refined by
M$^{\rm c}$Hardy \et (2005), who also included estimates based on host
galaxy gas kinematics and photoionisation modelling, to arrive at a
figure of $3-6 \times 10^6$~\Msun. Assuming a mass of $3\times
10^6$~\Msun\ the expected frequency of a HF QPO is $f_{\rm HFQPO} \sim
10^{-3}$~Hz. As shown in figure~\ref{fig:mcg6qpo}, around this
frequency range a QPO of strength $R\sim 2$ per cent cannot be ruled
out. The data are more sensitive to QPOs at high frequencies, but
even there a $R\sim 1$ per cent QPO would have been difficult to detect 
(the data are more sensitive to narrower QPOs, so if $Q>5$ the constraint
on $R$ is tighter).

\begin{figure}
\centering
\includegraphics[width=5.4 cm, angle=270]{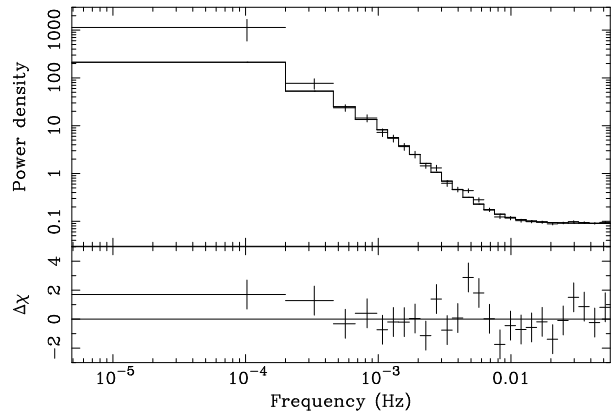}
\caption{
PSD of NGC 4051 calculated using the $0.2-10$~keV
light curve extracted from the \xmm\ EPIC pn.
The top panel shows the data (crosses) and model (histogram)
and the bottom panel shows the residuals.
The model is the bending power law model from M$^{\rm c}$Hardy \et
(2004), with the low frequency slope fixed to by $\alpha = 1.1$, as
measured from the \xte\ data.
Given the black hole mass of $0.5-1.0 \times 10^6$~\Msun\
a HF QPO might be expected around $f_0 \sim 1-3\times 10^{-3}$~Hz.
The positive excess in the residuals at around $5 \times
10^{-3}$~Hz is discussed in the text.
\label{fig:n4051psd}}
\end{figure}

The above analysis was repeated for the single-orbit \xmm\ long-looks
at the highly variable Seyfert 1 galaxies NGC 4051,
NGC 4395 and Mrk 766. These light curves have previously been discussed by M$^{\rm
  c}$Hardy \et (2004), Vaughan \et (2005) and Vaughan \& Fabian
(2003), respectively. 
The black hole mass for NGC 4051 is thought to lie in the range $\sim
0.5-1.0 \times 10^6$~\Msun\ (Shemmer \et 2003), which would
correspond to a HF QPO frequency of $\sim 1-3$~mHz.
For NGC 4395 the black hole mass is much lower, lying in the range
$10^4-10^5$~\Msun\ (Filippenko \& Ho 2003). This corresponds to a
HF QPO frequency of $\sim 30-300$~mHz.
For Mrk 766 the black hole mass is probably in the range $\sim
0.5-5\times 10^6$~\Msun\ based on its optical emission and absorption
line widths (using the stellar velocity dispersion $\sigma_{\ast}$
from Botte \et 2005, and the $M_{\rm BH}-\sigma_{\ast}$ relation of
Tremaine \et 2002, gives $\sim 3.5\times 10^6$~\Msun).
Therefore, for Mrk 766 a  HF QPO might be expected to appear around $\sim
1$~mHz.

In each case a broken power law provided an acceptable fit to the \xmm\
PSD with fitted parameters similar to
those of the previous  analyses (cited
above). Neither Mrk 766 not NGC 4395 show any QPO-like residuals.
For Mrk 766 the strength of a QPO
was constrained to be $R \ls 3$ per cent rms, at frequencies above a few mHz, 
while for NGC 4395 the limits were slightly
worse owing to the much lower mean count rate in that source
(which raises the Poisson noise level of the power spectrum).

The binned PSD for NGC 4051 was reasonably well fitted by a broken
power law ($\chi^2 = 31.85$ for $21$ dof).  Interestingly, the
data/model residuals showed a positive excess at $\approx
5$~mHz which could indicate a HF QPO.  After adding a Lorentzian (with
$Q=5$) to the model to fit the residual feature, the fit improved
dramatically, from $\chi^2 = 31.85$ to $\chi^2 = 14.76$ (for the
addition of two free parameters, centroid frequency and
normalisation).  However, it may be that these residuals are caused by
the use of an inappropriate continuum model rather than a genuine QPO.

M$^{\rm c}$Hardy \et
(2004) performed a simultaneous analysis of this \xmm\ observation and
long-term \xte\ monitoring data and found a bending power law model
gave the best fit to the broad band power spectrum. The low frequency
slope, constrained by the \xte\ data, was $\alpha_{\rm low} = 1.1$.
Thus, in order to better model the continuum power spectrum a
bending power law was fitted, with a low frequency slope fixed at
$1.1$. 
This gave $\chi^2 = 30.51$ for $22$ dof, a slight improvement
over the broken power law. Including an additional Lorentzian
QPO in the model, at $f_0 = 5 \times 10^{-3}$~Hz with a fixed width
($Q=5$), improved the fit to $\chi^2 = 17.44$ for $21$ dof. 
The strength of the Lorentzian was $R \approx 2.3$ per cent, which 
is plausible for a HF QPO. An $F$-test yields a probability of
only $p=7 \times 10^{-4}$ for this improvement\footnote{
As noted by Protassov \et (2002), the $F$-test is often 
inappropriate for calculating the significance of additive 
features such as QPOs, due to the null values of the additional 
parameters lying on the boundary of the possible parameter space. 
In the present context the only additional free
parameter was the Lorentzian normalisation, which was allowed 
to be positive or negative, hence the null value (zero) was
formally not on the boundary. The probability was then adjusted
for the number of possible trials (frequencies searched).
}. However, this 
does not account for the number of frequencies searched. 
A lower limit on this is the number of bins, in this case $N=26$.
A lower limit on the `global' probability for the QPO candidate
is therefore $p_N = 0.018$, using $p_N = 1-(1-p)^N$. Taken on its
own this indicates a possible QPO detection
at $\sim 98$ per cent confidence. One should, however, also 
account for confirmation bias: this is the best QPO candidate
found from an analysis of four datasets. The significance
should therefore be more accurately considered as $\sim 93$ 
per cent, not strong enough to be
considered a confident detection of a HF QPO. However, this
does illustrate that QPOs with strengths of order $2$ per cent rms
are only just below the sensitivity of the available data.
The analysis of these data by M$^{\rm c}$Hardy \et (2004) used much
coarser frequency binning and therefore therefore was not
sensitive to this fairly narrow feature. But the
presence of a QPO does not substantially affect the results of 
that analysis. Even after 
allowing for a Lorentzian HF QPO in the \xmm\ data, the bend frequency
was $f_{\rm b} = 1.2\pm0.4 \times 10^{-3}$~Hz, consistent with 
the value of $0.8_{-0.3}^{+0.4} \times 10^{-3}$~Hz 
measured by M$^{\rm c}$Hardy \et
(2004) using the combined \xte\ and \xmm\ data.

\begin{figure}
\centering
\includegraphics[width=6.40 cm, angle=270]{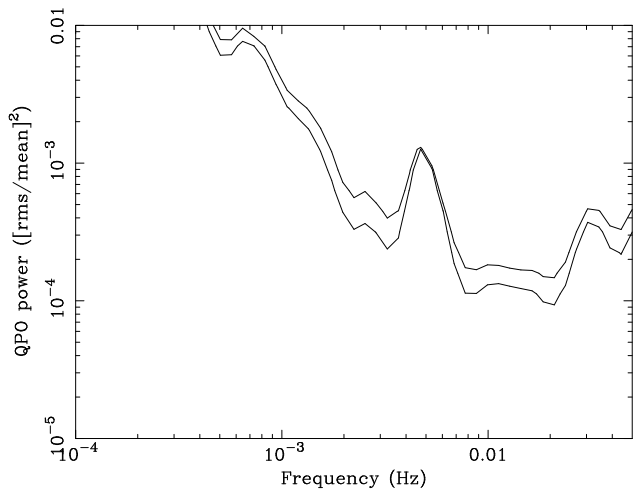}
\caption{
Limits on the strength of any HF QPOs in NGC 4051
based on the \xmm\ PSD (see Fig.~\ref{fig:n4051psd}).
The lines have same meaning as in Fig.~\ref{fig:mcg6qpo}.
\label{fig:n4051qpo}}
\end{figure}

Figure~\ref{fig:n4051qpo} shows the limits on the QPO strength
over the available frequency bandpass (calculated as for
figure~\ref{fig:mcg6qpo}, assuming no QPOs are present).

These power spectral fitting results suggest that the best available
high frequency data from \xmm\ can rule out strong ($R>$ few per
cent), narrow ($Q \gs 5$) HF QPOs at the expected
frequencies. However, the data are not quite sensitive enough to
detect weaker QPOs ($R \ls 2$ per cent) with confidence.


\section{Can we detect low frequency QPOs?}
\label{sect:lfqpo}

In the case of LF QPOs, the expected quasi-periods in Seyferts (assuming scaling from
a $\sim 1$~Hz QPO in a 10~\Msun\ GBH) are of the order of days--months (for
$10^{6}$--$10^{8}$~\Msun\ black holes) and therefore
require reasonably
well-sampled long-term X-ray monitoring to have any chance of being
detected. 
Such monitoring has been obtained by \xte\ since it began
operating in 1996. 
The All-Sky Monitor (ASM) on \xte\ is not
sensitive enough to accurately monitor Seyfert galaxies (see e.g.
Uttley \et 2002), and so pointed observations are required.
In order to maximise the
PSD frequency-bandwidth while reducing the required observing time,
the highest-quality \xte\ Seyfert monitoring light curves consist of
$\sim1$~ks duration pointed observations following several different
temporal sampling patterns which cover a range of sampling intervals, from a few hours
to several days  (e.g. see Markowitz \et 2003; M$^{\rm c}$Hardy \et
 2004, 2005).  
With these data, it has been possible to show that
all Seyfert PSDs measured to date are consistent with simple power laws
or (in the majority of cases) broken or bending power laws
(e.g. Uttley \et 2002; Markowitz \et 2003;  M$^{\rm c}$Hardy \et 2004,
2005).  Therefore, there is as yet no firm evidence for LF QPOs in
Seyfert X-ray light curves.  However, it is
important to understand whether  LF QPOs could even have been detected using
the best existing \xte\ monitoring data to date, if they are present 
in the light curves.

\begin{figure}
\centering
\includegraphics[width=6.0 cm, angle=270]{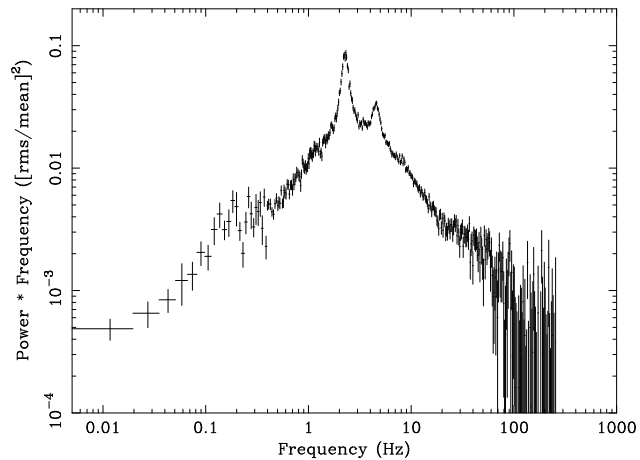}
\caption{
PSD of GRS 1915+105 calculated in the $2-13$~keV band using
the \xte\ observation of 1997 Feb 9.
Note the ordinate is in units of frequency$\times$power to better highlight the QPO.
\label{fig:grs1915psd}}
\end{figure}

Simulations were used to answer this question.
X-ray light curves were generated with
the same PSD shape as actually observed in GRS 1915+105, except scaled down
in frequency by the ratio of black hole masses.
For the input `true' PSD shape, the $2-13$~keV
PSD of the 1997 Feb 9 \xte\ observation of GRS~1915+105 was chosen,
as this showed a very prominent QPO at $f_0 \simeq2.26$~Hz
(with $R \simeq 15$ per cent, $Q \simeq 5$).
Fig.~\ref{fig:grs1915psd} shows the PSD of this observation, in units of
frequency$\times$power to better highlight the QPO.
The QPO is one of the strongest observed in GRS 1915+105 at this 
relatively high (for a LF QPO) frequency;
the combination of high fractional rms and relatively high frequency
(and hence more observed cycles)
should make it more easily detectable with the existing \xte\ monitoring sampling
patterns.  To obtain the power at each Fourier frequency in the simulation
the observed PSD was interpolated between adjacent bins, and above a
frequency of 15~Hz where the
observed GRS~1915+105 PSD becomes noisy, the PSD was
extrapolated using the best-fitting $7.5-15$~Hz
power law (index of $-2.4$).
Light curves with this PSD were simulated using the method of Timmer
\& K\"{o}nig (1995). 

Arguably the highest-quality long-term monitoring data obtained to date is for
the Narrow Line Seyfert~1 NGC~4051 (M$^{\rm c}$Hardy \et 2004), which
\xte\ has monitored for more than 8 years,
including observations every 2 days over the last 4 years, as well as a 
two month period of observations every 6 hours.  
The exact sampling pattern of the NGC~4051 monitoring
was used to simulate the Seyfert light curve.
In addition, a single `long-look' light curve, of duration
$130$~ks, with continuous sampling binned into $500$~s time intervals,
was included to
simulate an \xmm\ observation (which can constrain the PSD at 
high-frequencies), since many of the Seyferts with good \xte\ monitoring have also
been observed by \xmm.  Although the LF QPOs will
occur at lower frequencies than can be observed by a single \xmm\
orbit, the continuous sampling of the data 
constrains the high-frequency shape of the PSD, allowing aliasing
effects (which distort the \xte\ PSD) to be better constrained, and permitting
better detection of any sharp features above a broader PSD continuum.

The input PSD was rescaled by a factor $10^{-6}$ in frequency, corresponding
to a `typical' Seyfert black hole mass of $10^{7}$~\Msun\ 
(Peterson \et 2004), and from this a continuous light curve with $500$~s resolution
was simulated and re-sampled to match the sampling patterns described above.
In order to demonstrate the effects of the intrinsic stochastic character
of the light curve on detecting a QPO, the effects of photon counting 
statistics were neglected.  In any case, the effects of counting statistics on
the highest-quality AGN PSDs measured to date are negligible, except at the highest 
frequencies (comparable to the $500$~s sampling used here).
The {\sc psresp} Monte Carlo method of Uttley \et (2002) was then used
to fit models to the observed PSDs (see also
Markowitz \et 2003, M$^{\rm c}$Hardy \et 2004).  

\begin{figure}
\centering
\includegraphics[width=6.0 cm, angle=270]{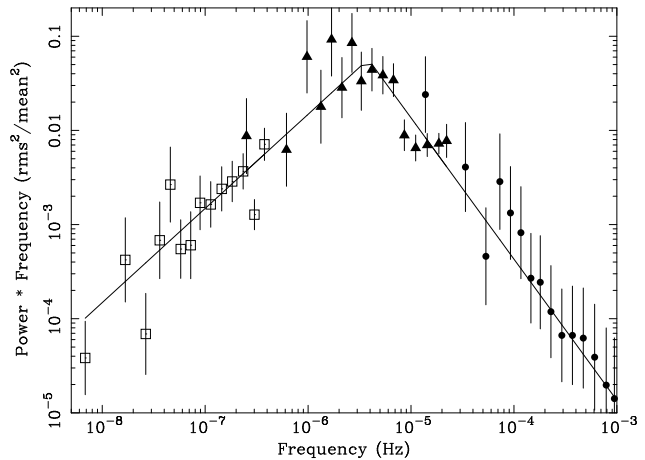}
\caption{
Simulated `unfolded' PSD of a Seyfert galaxy showing a strong LF QPO in the underlying
light curve, assuming sampling to match the best so far obtained for an AGN
(see M$^{\rm c}$Hardy \et 2004). 
The simulated data were generated using the observed PSD of GRS 1915+105 (see
Fig.~\ref{fig:grs1915psd}) with the frequencies
scaled as appropriate for a $M_{\rm BH}=10^7$~\Msun\ Seyfert galaxy (see text for details).  
The different data points mark the PSD contributions
from \xte\ long time-scale (8 years at varying sampling
intervals; squares) and intensive 6-hourly  
monitoring (filled triangles) and a $130$~ks continuous
({\it XMM-Newton}-like) long-look observation (filled circles).  The
strong QPO cannot be distinguished  
from a broken power law continuum model (solid line).
\label{fig:n4051psd-sim}}
\end{figure}

The simulated `observed' PSD is shown in Figure~\ref{fig:n4051psd-sim}, together with the
best-fitting broken power law continuum model, assuming a low-frequency slope which
was fixed at $0$ (i.e. similar to that observed at low frequencies in the
GRS~1915+105). 
To aid the comparison, the simulated Seyfert data is
plotted on the same relative scale as 
the GRS~1915+105 data (Fig.~\ref{fig:grs1915psd}), with the
frequencies scaled down by a factor of $10^{6}$. Note that this model
represents only a noise continuum spectrum and contains no QPOs.
The figure shows the `unfolded' PSD, i.e. data points
have been plotted against the best-fitting model in terms
of their positions relative to a model that has been distorted by
sampling effects. By analogy with 
plotting an X-ray spectrum which is unfolded through the instrumental response
with respect to an assumed spectral model, the effects of unfolding the PSD are relatively 
model-independent for the broad continuum model used here.  
The fitted model has a break at
$3.8\times 10^{-6}$~Hz (cf. the QPO frequency of $f_0 = 2.26\times 10^{-6}$~Hz) and
high-frequency slope of $-2.5$, and is formally acceptable ($40$ per
cent of simulated data sets with the same broken power law
model parameters showed worse fits, defined in terms of a $\chi^2$ statistic, than
the `observed' data\footnote{See Uttley \et (2002) for more details of the goodness-of-fit
calculation.}).  

Clearly the strong, sharp QPO that is present in the `true' PSD
(Fig.~\ref{fig:grs1915psd}) could not  be easily discerned in the PSD
of the simulated Seyfert data (Fig.~\ref{fig:n4051psd-sim}).  This
lack of sensitivity to the QPO occurs for two reasons. Firstly, the
QPO is spread by the redistribution of power (aliasing) caused by the
intermittent sampling and finite duration of the intensive 6-hourly
monitoring (e.g. see van der Klis 1989a and Uttley \et 2002 for
discussion of aliasing).   Secondly, the error bars on the data are
large in the frequency range of interest, because the 2-month duration
of the 6-hourly monitoring  did not span many cycles of the
variability at those frequencies.  Even extending the high-frequency
end of  the low-frequency (long-term monitoring) component of the
total PSD to the maximum (Nyquist) frequency ($\simeq5.8\times
10^{-6}$~Hz) of the 2-day sampled monitoring (which lasted 4 years)
does not improve the sensitivity to the QPO, because aliasing effects
are substantially increased.

The test was repeated using different frequency scalings for the QPO,
i.e. to allow for different black hole masses and the range of
observed QPO frequencies in  GRS~1915+105, but in no case was the QPO
clearly distinguishable from a broader broken power law PSD.  At lower
frequencies, there were too few QPO cycles and error bars are larger,
while  at higher frequencies the sampling is not sufficiently dense
and aliasing effects wash out the  sharp QPO peak.   Clearly, even
this strong QPO could not have been detected if present  in the best
existing Seyfert monitoring data.


\section{Discussion and conclusions}
\label{sect:disco}

\subsection{Where are the AGN QPOs?}

The similarities between the variability properties of AGN and GBHs
lead us to expect, by analogy, that some AGN will show QPOs similar to
those observed in GBHs but at much lower frequencies.  This
expectation does depend on the physical mechanisms for QPO generation being
similar in AGN and GBHs, but there is no clear theoretical reason to
expect otherwise.  Therefore, given the limits on the strength of any
HF QPOs and simulations of LF QPOs presented above, it is
possible to answer why the best, presently available (\xmm\ and \xte)
data have not revealed the expected QPOs.

\subsubsection{High-frequency QPOs}

In the case of HF QPOs, the most
obvious reason why they have not yet been observed is that the
current observations are not sensitive enough to detect them, even if they are
present.  

Even for the most variable AGN so far observed with \xmm\ -- which
probably have low black hole masses and hence the HF QPOs should be
well-sampled by \xmm\ (with typically  $\gs 100$ cycles) -- the
observations are only sensitive to QPO strengths of $R \gs 2$~per cent.
Given that many detected HF QPOs in GBHs are less coherent than the
assumed $Q=5$ {\it and} that HF QPO strengths greater than  a few per
cent are not very common, it is perhaps not surprising that none were detected in the
Seyfert light curves.  

In order to detect a QPO in a time series, the
series must be sufficiently long that is spans
many cycles of the quasi-period.
The number depends on the width of the QPO, but a general
rule-of-thumb, based on the analysis of section~\ref{sect:hfqpo}, is
that considerably more than $100$ quasi-periods must be observed before 
the errors on the PSD are small enough to distinguish weak HF QPOs
from sampling fluctuations in the underlying aperiodic noise spectrum.
(The signal must be sampled at more than twice the peak
frequency of the QPO otherwise the QPO falls above the Nyquist
frequency of the PSD, which is not directly measured and
is aliased onto the wrong frequencies.)
It would therefore be sensible to treat with caution any claim of
quasi-periodic variations in AGN light curves that is based on only a
small number of cycles of the claimed QPO peak frequency.

HF QPOs are rather rare even in GBH light curves; they
have been detected in $<10$~per cent of observations in less than
$1/3$ of GBH candidates (see tables in Remillard \et 2002a,b;  McClintock
\& Remillard 2005).  Although not often present, HF QPOs in GBHs
appear to be confined to sources in the `very-high' state, an
apparently distinct accretion state which shows X-ray spectra with
a strong, steep power law component and in some cases a strong thermal
component also (McClintock \& Remillard 2005; van der Klis 2005). If
the AGN observed to date are not in this state (which most likely
corresponds to high accretion rates), perhaps HF QPOs would not be
expected.

Another factor which may reduce the likelihood of observing HF QPOs in
AGN light curves, at least in {\it XMM-Newton} data, is the fact that
the HF QPO strength increases towards higher X-ray
energies (Strohmayer 2001; Remillard 2002a) whereas {\it XMM-Newton} is most
sensitive in the soft X-ray band ($<$2~keV).  Unfortunately, most GBHs
are quite  heavily obscured in soft X-rays by large Galactic column
densities ($N_{\rm H} > 10^{22}$~cm$^{-2}$; McClintock \& Remillard 2005)  so that
it is not possible to say for certain what their timing properties are
at low energies, but simple extrapolation from the energy-dependent
behaviour above 2~keV would suggest that their soft X-ray QPO
strengths are even weaker than are observed at higher energies.

Of course, absence of evidence is not evidence of absence. It is not
possible to rule out the possibility that weak ($R \ls 2$ per cent) and/or
low-coherence ($Q \ls 5$) QPOs might exist in AGN, or that  transient
QPOs might exist in AGN that are not easily detected in GBHs
(especially if their frequencies are variable).   However, since the
ubiquitous aperiodic variability can easily mimic  apparent
quasi-periodicities one must be careful in assessing whether these
apparent signals are significant and represent QPOs in any meaningful
sense.  Any claimed QPOs must be coherent and long-lasting enough that
they form peaks that lie significantly above the expected statistical
variations in the continuum level of the PSD (e.g. see
Benlloch \et 2001, Vaughan 2005).  

\subsubsection{Low-frequency QPOs}

Despite the fact that LF QPOs are typically much stronger than the HF
QPOs, the detection of LF QPOs is made difficult by the relatively
sparse sampling of the available monitoring light curves, which
although excellent for constraining the shapes of broad continua, are
not intensively sampled for long enough to detect even strong QPOs.
However, despite this lack of sensitivity it is interesting to note
that the broad-band shape of the simulated PSD shown in
Figure~\ref{fig:n4051psd-sim} is quite distinctive, showing a strong
deficit in low-frequency power (i.e.  the power is clearly
band-limited) and is similar to the broad continuum shape of the input
PSD.  
Therefore, with data like Fig.~\ref{fig:n4051psd-sim}, a PSD similar
to that of GRS~1915+105 on 1996 Feb 9, which 
shows a strong QPO, might be mistaken for a PSD like that of GBHs in
the low/hard state, which have a similar  band-limited PSD shape
(e.g. McClintock \& Remillard 2005; van der Klis 2005).  Currently,
two AGN show evidence for a band-limited PSD shape with a
low-frequency cut-off in power: the broad line Seyfert NGC~3783
(Markowitz \et 2003) and the NLS1 Ark~564 (Pounds \et 2001; Papadakis
\et 2002; Markowitz \et 2003).  These AGN may be good candidates for
future LF QPO searches with  longer periods of intensive sampling.

Of the other AGN, NGC~4051 and MCG$-$6-30-15 show the highest-quality
PSDs, with relatively short characteristic time-scales so that any low-frequency cut-offs similar
to that seen in Figure~\ref{fig:n4051psd-sim} should be easily observed.  In fact, 
these PSDs show no evidence for low-frequency cut-offs, instead showing
$1/f$ flickering
extending down to the lowest observed frequencies ($10^{-8}$~Hz), a fact which makes
them appear similar to the PSDs of certain GBHs (e.g. Cyg~X-1) in the high/soft state.

\subsection{How might AGN QPOs be detected in the future?}

\subsubsection{High-frequency QPOs}

The two characteristics of the observations that  most strongly affect
the sensitivity to  QPOs are count rate (more specifically, the
signal-to-noise of the source fluxes) and  the length of the
observation. (If the sampling is not contiguous then the sampling
pattern will also be an important factor.)  The frequency of the QPO
relative to the  other components present in the spectrum determines
which of these two is dominant.  If the Poisson noise is the
strongest continuum component at the QPO frequency (i.e. at $f_0$ the
Poisson noise level is above the intrinsic red noise continuum power)
then the sensitivity to QPOs is a strong function of the count
rate. The Poisson noise level is inversely proportional to the count
rate, meaning a larger telescope collecting area will drastically
reduce the Poisson noise level,  revealing weaker QPOs. As discussed
by van der Klis (1989b), the significance (in units of sigma) at
which a QPO can be detected against a Poisson noise background
is given approximately by 
$n_{\sigma} = \frac{1}{2} I R^2 (T / \Delta f)^{1/2}$, where
$I$ is the mean count rate, $R$ is the QPO strength, 
$T$ is the observing time and $\Delta f$ is the width of the QPO. 

If however the intrinsic red noise is greater than the Poisson noise
at the QPO frequency  (i.e. the dominant source of noise is the
aperiodic variability of the source), then reducing the Poisson noise
level with a larger telescope will have little effect. (In essence the
signal-to-noise is already good enough that  Poisson noise is not the
main contaminant at the QPO frequency.) In this case the most
important factor is the number of cycles observed, i.e. the duration
of the observation, or  the number of repeat observations.  The errors
on the PSD (binned over a fixed frequency range) scale as
$\sqrt{T}$. A large increase in the total amount of data will reduce
the sampling fluctuations in the measured red noise continuum,
allowing weak QPOs to be detected.

\begin{figure}
\centering
\includegraphics[width=5.0 cm, angle=270]{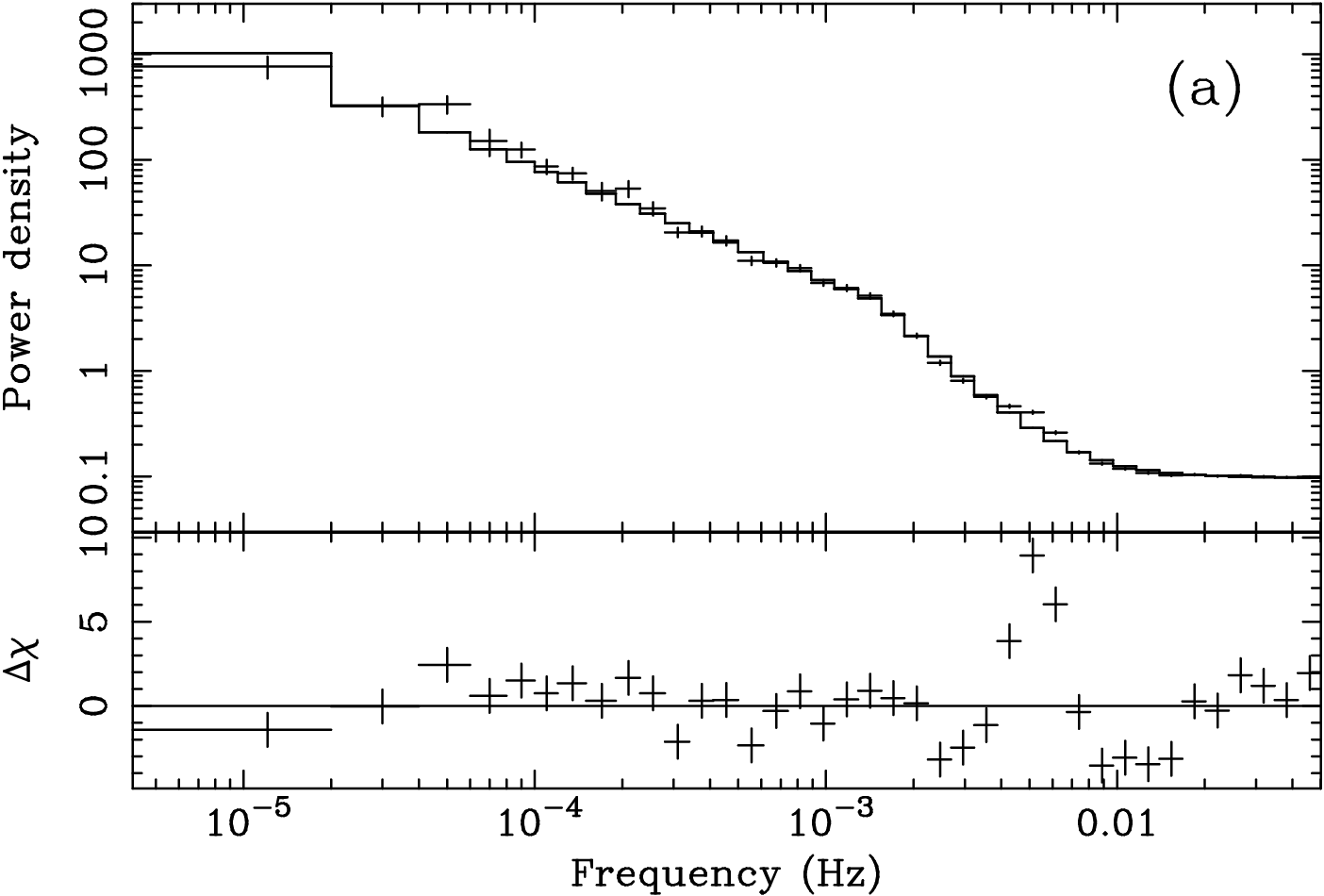}
\\
\vspace{0.5 cm}
\includegraphics[width=5.0 cm, angle=270]{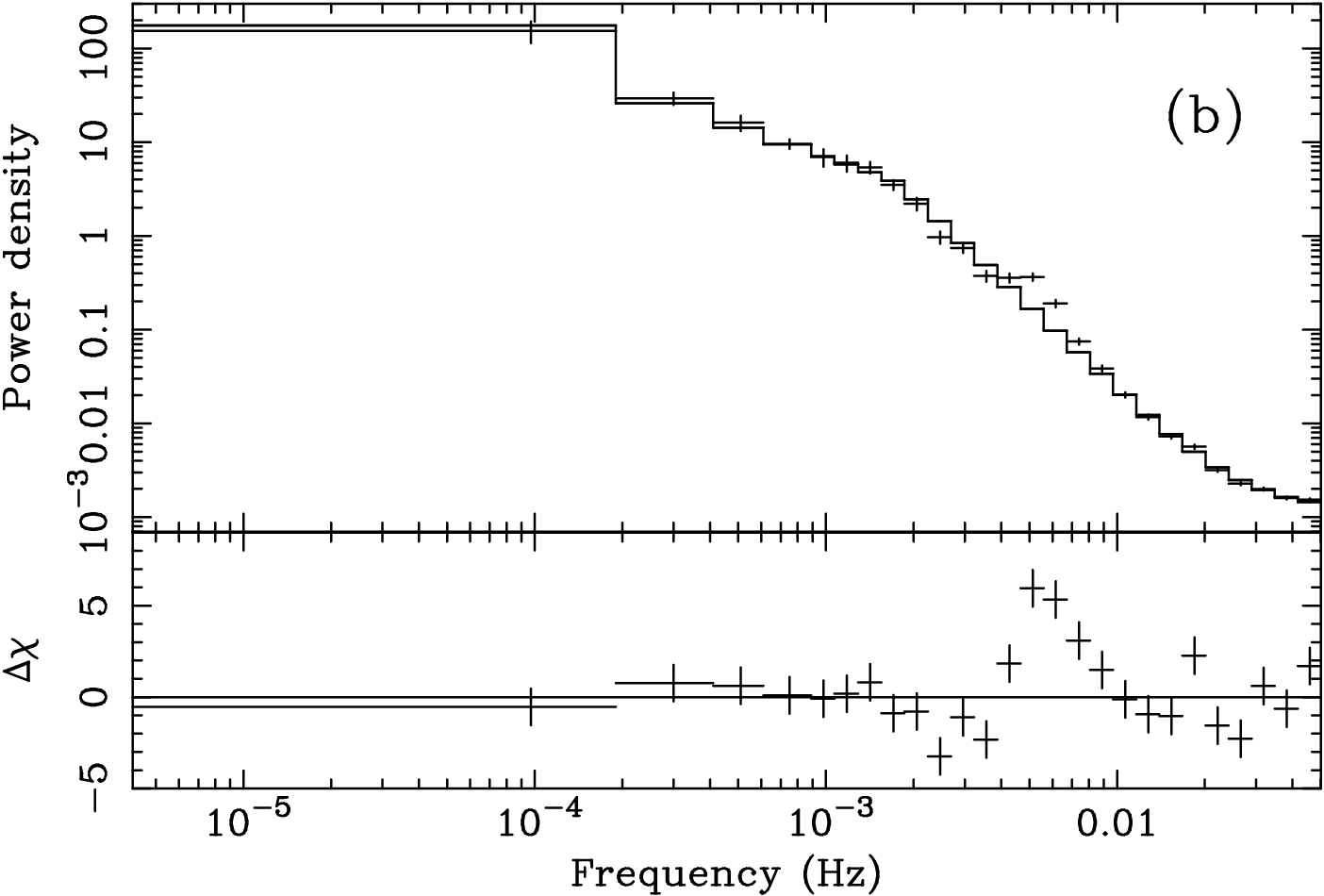}
\caption{
Simulations of the PSD of NGC 4051 assuming a HF QPO
at $f_0 = 5 \times 10^{-3}$~Hz (with $Q=5$ and $R=0.02$).
Panel $a$ shows the PSD from $10$ orbits of \xmm\ data,
each containing $120$~ks of continuous sampling.
Panel $b$ shows the PSD from a single $120$~ks exposure
with the \xeus.
In each case the simulated PSD has been fitted with a 
broken power law continuum model (plus Poisson noise) 
to illustrate the QPO.
}
\label{fig:xeus}
\end{figure}

Figure~\ref{fig:xeus}$a$ and $b$ illustrate these two effects. The
figures show simulated PSDs for  NGC 4051 generated using a broken
power law continuum and a QPO at $f_0 = 5 \times 10^{-3}$~Hz (with
$Q=5, R=0.025$), and including Poisson noise at the appropriate
level. This QPO model was chosen to correspond to the
feature in the NGC 4051 data 
discussed in section~\ref{sect:hfqpo}, which  is plausible for a
HF QPO but is not a  significant detection based on the existing data.
The QPO is at a frequency where the power density in the intrinsic red noise 
is comparable to the Poisson noise level in the \xmm\ observation.
Panel $a$ shows the PSD derived from $10$ orbits of \xmm\ data, with a
combined exposure time of $1.2$~Ms.  Panel $b$ shows the PSD derived from
a simulation of a $120$~ks observation with the Wide-Field Imager
(WFI) camera on-board ESA's proposed next generation X-ray
telescope, \xeus, with a collecting area of order $10$~m$^2$ (Parmar
\et 2001)\footnote{
The \xeus\ WFI count rate was estimated by fitting the \xmm\
EPIC NP spectrum with a simple model and convolving this with the
publically available \xeus\ response matrices available from
{\tt ftp://www.rssd.esa.int/pub/XEUS/RESPONSE/}
}.  
The errors in the continuum PSD are reduced using the 
several \xmm\ orbits, but the Poisson noise level stays 
the same, whereas the Poisson noise level is greatly reduced
using the same observation length but a much larger
collecting area.
It is clear from the figures that both of these observations are
capable of detecting the QPO, with very high confidence, but the
many-orbit \xmm\ observation provides the best detection, despite the
Poisson noise level being a factor $\sim 100$ higher than in the
\xeus\ data.

\subsubsection{Low frequency QPOs}

\begin{figure*}
\hbox{
\hspace{0.1 cm}
\includegraphics[width=4.0 cm, angle=270]{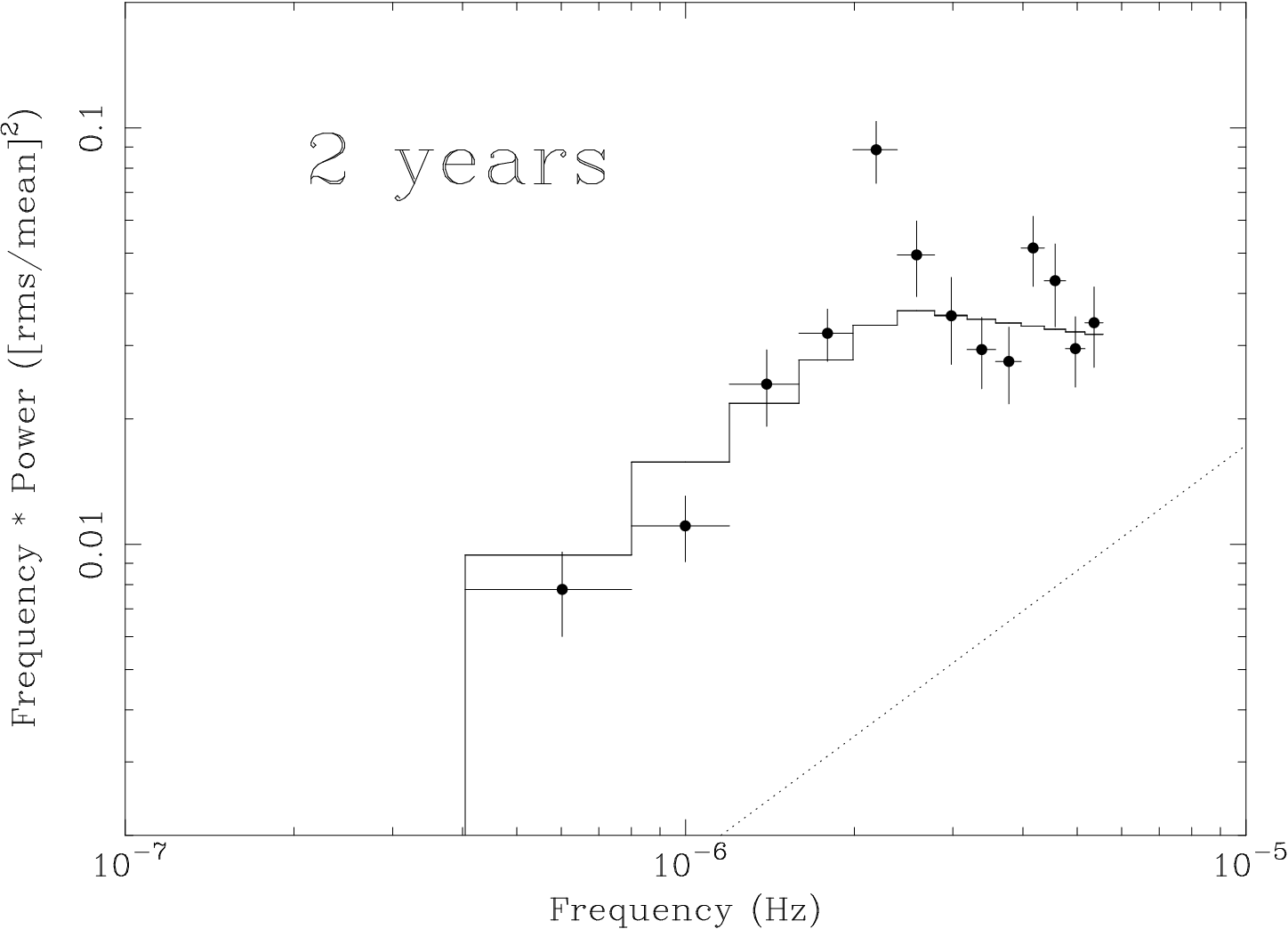}
\hspace{0.1 cm}
\includegraphics[width=4.0 cm, angle=270]{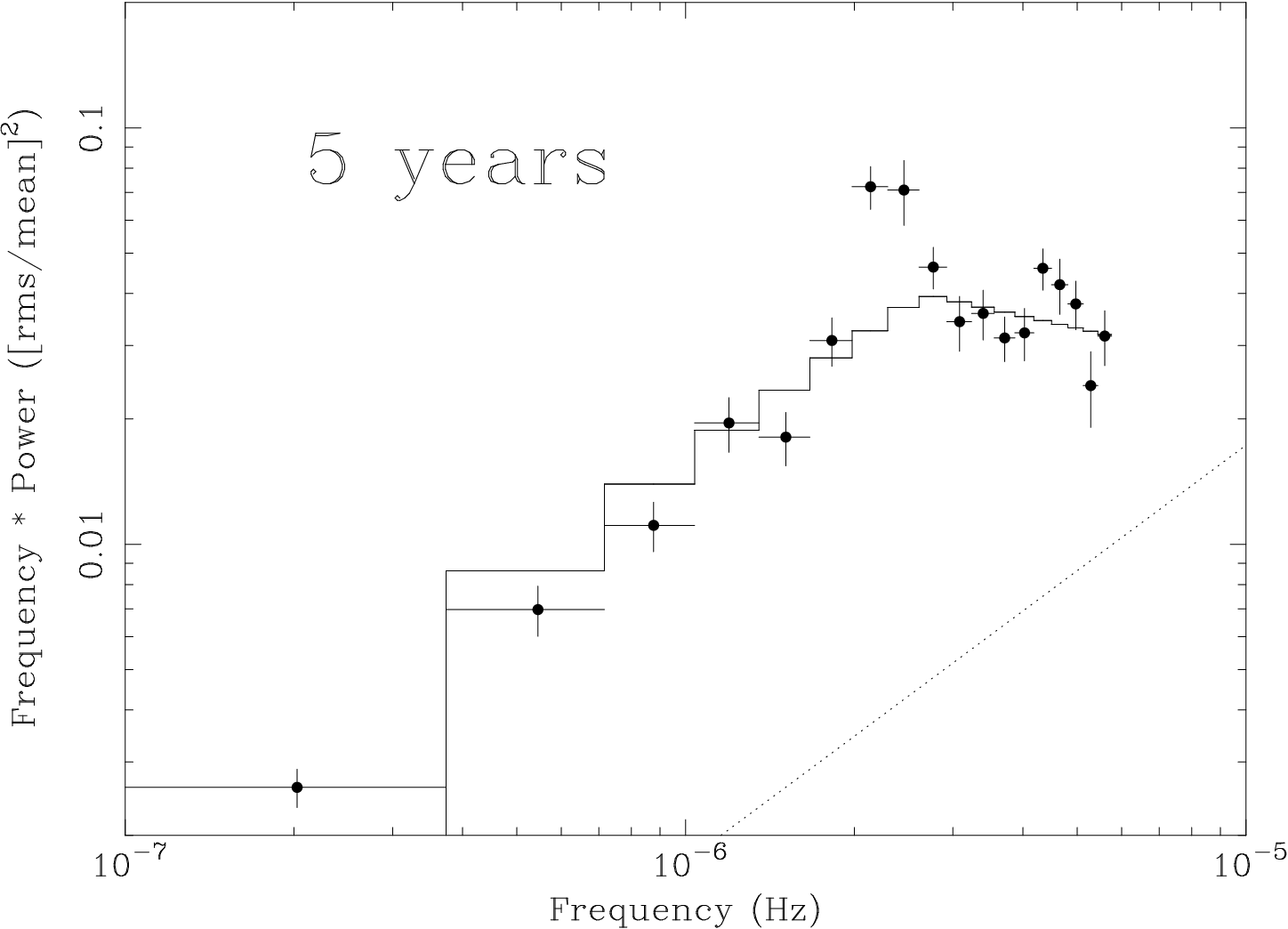}
\hspace{0.1 cm}
\includegraphics[width=4.0 cm, angle=270]{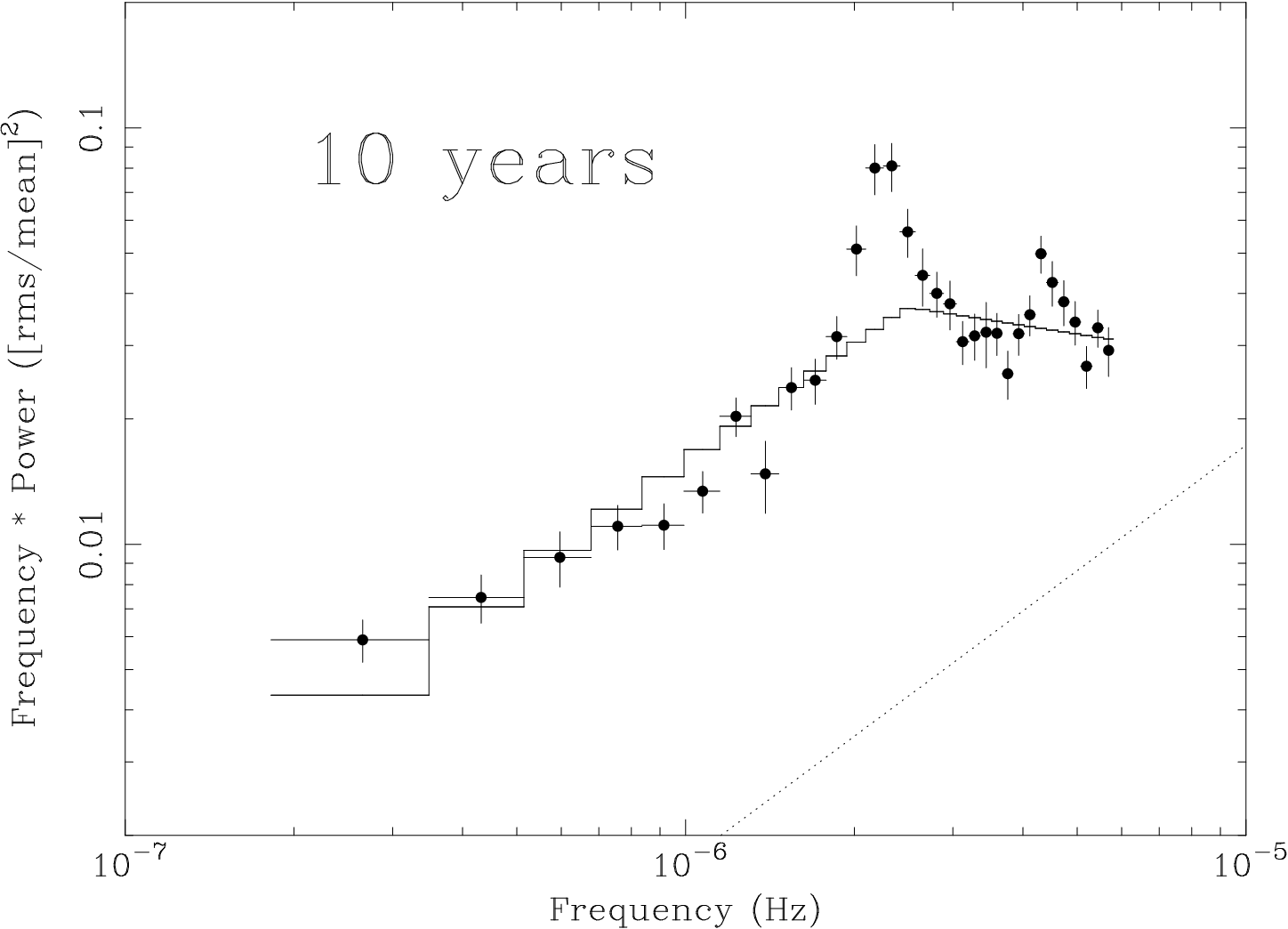}
}
\caption{
Simulated PSD from 2, 5 and 10 years of continuous observations from an
all-sky monitor, assuming the same underlying PSD shape as used in
Figure~\ref{fig:n4051psd-sim} and continuous daily sampling.  
Note the ordinate is in units of frequency$\times$power to better highlight the QPO.
The
solid line shows the best-fitting broken power law PSD.  Photon counting
statistics are not included in the simulation, but are negligible on the
relevant time-scales, as shown by the dotted line which
shows the expected noise level for a source detected at 10-sigma in one day,
consistent with the sensitivity of proposed ASMs and wide-field monitors.  
Even without any high-frequency data, the QPO and its harmonic are easily detected
(see text for further details).
\label{fig:asmqpo}}
\end{figure*}

In order to detect LF QPOs, observed AGN light curves must satisfy
two criteria: they must have a long duration, so that $>100$ QPO
cycles can be sampled, and they must have good sampling, so that sharp
QPO features are not smeared out by aliasing effects.  Enormous amounts
of on-source observing time are required to satisfy these criteria,
so that LF QPO detection programmes will be prohibitively expensive
for pointed observations.  Instead, the detection of LF QPOs in AGN
can be achieved relatively cheaply by future, sensitive X-ray All-Sky
Monitors (ASMs).

The power of long-baseline, continuous observations
of AGN with ASMs was demonstrated using simulations.
Artificial time series were generated using the same input PSD
as used in Section~\ref{sect:lfqpo} (i.e. assuming a $10^{7}$~\Msun\ black 
hole), sampled on 1-d time-scales over baselines of 2 years, 5 years and
10 years. Since the data are continuously sampled on day time-scales, 
aliasing effects are 
negligible, the data can be binned in frequency and fitted in the same way as for
continuous long-look observations (see Section~\ref{sect:hfqpo}).  
In order to test the significance of the QPO detections,
a broken power-law was first fitted (with all parameters left free) as
a null hypothesis, to which  Lorentzian QPO features were added
to model the QPO fundamental and harmonic.
The $F$-test was used to determine if the QPO features are required 
by the data.  The simulated PSDs (and best-fitting broken power-law 
models) are shown in Figure~\ref{fig:asmqpo}.

For the PSD measured over a 2-year baseline,
a Lorentzian QPO at $f_0 = (2.2\pm0.14)\times 10^{-6}$~Hz was required by the
data at 
99.5~per~cent confidence, from a na\"{\i}ve application of the $F$-test 
($\Delta \chi^{2}=-19.1$ for 2 additional free parameters, FWHM and 
normalisation). The number of frequencies searched should be accounted
for separately, by adjusting the confidence level to account for the
number of trials (e.g. see Vaughan 2005). Since the QPO feature contributes
primarily to 2 out of 14 frequency bins (see Figure~\ref{fig:asmqpo}), the 
actual confidence limit, with no a priori expectation of a QPO at that 
frequency is more like $0.995^{7}\sim97$~per cent (i.e. a 2-$\sigma$ detection).

Extending the baseline of the simulated light curve to 5 years, 
the QPO at $f_0 = (2.28\pm0.07)\times 10^{-6}$~Hz
was detected at $>99.9$~per cent confidence (i.e. $>3$-$\sigma$)
after accounting for the number of bins sampled, although the harmonic
could not be significantly detected. Finally,
for the 10-year baseline simulation the
QPO fundamental [at $f_0 = (2.23\pm0.04)\times 10^{-6}$~Hz] was required at
$>99.9999$~per~cent confidence (i.e. $\sim 5$-$\sigma$), {\it and} the
harmonic is detected at twice the fundamental frequency at $>97$~per~cent
confidence.  Considering the number of sources
which will be monitored by a sensitive ASM
(and hence the increased likelihood of `false alarm' detections),
we should require a QPO signal
to be detected at the 3-$\sigma$ level or better to be considered
robust. 
Therefore continuous monitoring of sources for 5 years is probably
required to significantly detect the presence of the anticipated LF QPOs, while
monitoring for 10 years can reveal additional information such as the
presence of higher harmonics.  

Note that the effects of photon counting statistics were not included in 
the simulations,
which were intended to show the minimum temporal baselines required for 
detection of the QPO signals.  However, the effects of 
counting statistics on LF QPO detection should be negligible for future ASMs.  
To demonstrate this fact, the PSDs plotted in Figure~\ref{fig:asmqpo} 
show the expected noise levels for sources which are detected at the 
10-$\sigma$ level in one day, which corresponds to the 
sensitivity to bright (few $10^{-11}$~erg~cm$^{2}$~s$^{-1}$) AGN for the
proposed  {\it Lobster} soft X-ray ASM (Priedhorsky, Peele \& Nugent 1996; Fraser \et
2002) and {\it EXIST} hard X-ray ASM (Grindlay 2004).  Since these instruments can integrate
flux on longer time-scales than one day and thus observe fainter AGN,
the sample of AGN
where LF QPOs (if they exist) could be detected will number in the hundreds, 
provided the temporal baselines for monitoring are long enough to discern
LF QPOs from broken power-law PSDs.

In all the simulations, LF QPO frequencies can
be very accurately constrained (to within a few per cent).  If additional 
calibrations can be found to account for the range of possible LF QPO 
frequencies\footnote{In GBHs there is a tight correlation 
between LF QPO frequency and X-ray spectral index (Vignarca 
\et 2003, and see Titarchuk \& Fiorito 2004 for a theoretical 
interpretation).  Therefore the X-ray spectral shape may also be a
good calibrator of QPO frequency in AGN.} then the detection of LF 
QPOs in AGN PSDs could afford highly accurate 
estimates of the AGN black hole mass.

Finally, it is worth considering the types of AGN that might be expected to 
show LF QPOs.  Since strong LF QPOs are not observed in GBHs in the 
high/soft state, but are observed sometimes in the low/hard state and 
rather frequently in the very high state (e.g. McClintock \& Remillard 
2005), it is likely that they will be observed in AGN in the equivalent 
states (if they exist).  Since GBHs in the low/hard and very high states 
can show relatively strong jet emission, radio-loud AGN (and also 
low-luminosity AGN which are relatively radio-loud) may be the best 
candidates for these states in AGN [e.g. as recently noted by
Jester (2005) and Nagar \et (2005)] and hence may be the most promising
sources to detect LF QPOs.

\subsection{Concluding remarks}

The above discussion may read as if X-ray timing
studies of AGN will always be inferior to those of GBHs. 
However, in some respects AGN can provide data that are superior to
those from the much faster and brighter GBHs.
A  Seyfert galaxy with $M_{\rm BH} = 10^6$~\Msun\ is expected to show
characteristic variability timescales that are $10^5$ longer than
those of a
$10$~\Msun\ GBH, but 
the X-ray flux of a typical Seyfert galaxy is only a factor $\sim
10^{3}$ smaller than that of a typical GBH. 
Therefore, a Seyfert galaxy
can provide $\sim 10^2$ more photons per characteristic timescale.
This much higher counts/timescale rate means that if QPOs can be
detected in AGN it should be possible to directly observe the
wave-form (in the time domain) with far greater precision than is 
possible for GBHs. This could in principle allow the details of the
X-ray emitting region to be constrained in a manner analogous to 
modelling of neutron star hot spots (e.g. Weinberg \et 2001).

The existing \xmm\ data for the best studied Seyferts
(see e.g. Fig~\ref{fig:mcg6psd} and \ref{fig:n4051psd}) 
already surpass the best GBH data in terms of measuring the
PSD at the very highest frequencies.
The \xmm\ long-looks constrain the intrinsic red noise PSDs up to 
frequencies at least as high as $\sim 3$~mHz, 
which is equivalent to the light-crossing time of 
$20$ gravitational radii for a $M_{\rm BH} \sim 3 \times 10^6$~\Msun\
Seyfert galaxy such as MCG$-$6-30-15 ($t_{\rm lc} = 20r_g / c = 20GM_{\rm
  BH}/c^3$).  For a GBH with $M_{\rm BH} \approx 10$~\Msun\ the frequency
corresponding to the light-crossing time of $20$ gravitational radii
is $\sim 1$~kHz, which remains out of reach with current observations
(see e.g. Revnivtsev \et 2000). In this way, X-ray timing studies of AGN and GBHs
do offer complimentary views of accretion on to black holes.


\section*{Acknowledgements}

This thank an anonymous referee for a thorough and constructive
referees report.
Based on observations obtained with \xmm, an ESA science mission with
instruments and contributions directly funded by ESA Member States and
the USA (NASA). SV thanks the PPARC for financial support.  PU acknowledges
support from the National Research Council Research Associateships program.


\bsp
\label{lastpage}

\end{document}